\documentclass[12pt]{article}
\usepackage{amssymb} 
\usepackage{amsmath,amsthm}

\begin{document}
\font\frak=eufm10 scaled\magstep1
\font\fak=eufm10 scaled\magstep2
\font\fk=eufm10 scaled\magstep3
\font\black=msbm10 scaled\magstep1
\font\bigblack=msbm10 scaled\magstep 2
\font\bbigblack=msbm10 scaled\magstep3
\font\scriptfrak=eufm10
\font\tenfrak=eufm10
\font\tenblack=msbm10

%A continuacion definimos los comandos para utilizar los
%fuentes en modo matematico

%Operadores especiales, abrev. matematicas
\def\biggoth #1{\hbox{{\fak #1}}}
\def\bbiggoth #1{\hbox{{\fk #1}}}
\def\sp #1{{{\cal #1}}}
\def\goth #1{\hbox{{\frak #1}}}
\def\scriptgoth #1{\hbox{{\scriptfrak #1}}}
\def\smallgoth #1{\hbox{{\tenfrak #1}}}
\def\smallfield #1{\hbox{{\tenblack #1}}}
\def\field #1{\hbox{{\black #1}}}
\def\bigfield #1{\hbox{{\bigblack #1}}}
\def\bbigfield #1{\hbox{{\bbigblack #1}}}
\def\Bbb #1{\hbox{{\black #1}}}
\def\v #1{\vert #1\vert}             %Para denotar elgrado de #1 
\def\ord#1{\vert #1\vert} 
\def\m #1 #2{(-1)^{{\v #1} {\v #2}}} %Para denotar el signo (-1)^...
\def\lie #1{{\sp L_{\!#1}}}               %%Lie derivative
\def\pd#1#2{\frac{\partial#1}{\partial#2}}
\def\pois#1#2{\{#1,#2\}}
 %  un parentesis de Poisson {f,g}
\def\set#1{\{\,#1\,\}}             %  notacion para conjuntos
\def\<#1>{\langle#1\rangle}        %  una forma bilineal <x,a>
\def\>#1{{\bf #1}}                %  notacion para vectores 
\def\f(#1,#2){\frac{#1}{#2}}
\def\cociente #1#2{\frac{#1}{#2}}
\def\braket#1#2{\langle#1\mathbin\vert#2\rangle} %% <w|z>
\def\brakt#1#2{\langle#1\mathbin,#2\rangle}           %% <w,z>
\def\dd#1{\frac{\partial}{\partial#1}} 
\def\bra #1{{\langle #1 |}}
\def\ket #1{{| #1 \rangle }}
\def\ddt#1{\frac{d #1}{dt}}
\def\dt2#1{\frac{d^2 #1}{dt^2}}
\def\matriz#1#2{\left( \begin{array}{#1} #2 \end{array}\right) }
\def\Eq#1{{\begin{equation} #1 \end{equation}}}
\def\uno{\relax{\rm 1\kern-.28 em I}}

%%Abreviaturas de simbolos
\def\bw{{\bigwedge}}      %%from Marmo diff geom
\def\hut{{\scriptstyle \land}}            %%from Marmo diff geom
\def\dg{{\goth g^*}}                                                                                                            %%dual of the Lie algebra
\def\Cdg{{C^\infty (\goth g^*)}}
\def\poi{\{\:,\}}                           %parentesis de Poisson {,}
\def\qw{\hat\omega}                %  omega con sombrero
\def\FL{{\sp F}L}                 %  abreviatura para Transf. Legendre
\def\hFL{\widehat{{\sp F}L}}      %  abreviat. para Transf. Legendreext. 
\def\XHMw{\goth X_H(M,\omega)} 
\def\XLHMw{\goth X_{LH}(M,\omega)}                  
\def\ea{\varepsilon_a}
\def\ep{\varepsilon}
\def\mitad{\frac{1}{2}}
\def\x{\times}  
\def\cinf{C^\infty} 
\def\forms{\bigwedge}                 %  formas
\def\onda{\tilde}
\def\orb{{\sp O}}

%%letras griegas
\def\a{\alpha}
\def\g{{\gamma }}                  %  gama 
\def\G{{\Gamma}}	
\def\La{\Lambda}                   %  lambda
\def\la{\lambda}                   %  Lambda 
\def\w{\omega}                     %  una forma simplectica
\def\W{{\Omega}}                   %  Omega
\def\ltimes{\bowtie} 
             
%%letras cal, Bbb, goticas, etc.
\def\roc{{\tilde{\cal R}}}                       %%from Marmo diff geom
\def\cl{{\cal L}}                               %%from Marmo diff geom
\def\V{{\sp V}}                                 %espacio de velocidades
\def\F{{\sp F}}
\def\cv{{{\goth X}}}                    %  un campo vectorial
\def\LG{\goth g}
\def\LH{\goth h}
\def\X{{{\goth X}}}                     %  un campovectorial 
\def\R{{\hbox{{\field R}}}}             %%real numbers (Pepin)
\def\big R{{\hbox{{\bigfield R}}}}
\def\bbig R{{\hbox{{\bbigfield R}}}}
\def\C{{\hbox{{\field C}}}}         %%complex numbers (Pepin)
\def\Z{{\hbox{{\field Z}}}}             %%real numbers (Pepin)
\def\N{{\hbox{{\field N}}}}         %%complex numbers (Pepin)
%\def\small C{{\hbox{{\smallfield C}}}}         %%smallcomplex numbers (Pepin)

%%notaciones rm en modo math
\def\ima{\hbox{{\rm Im}}}                               %%Image of a map
\def\dim{\hbox{{\rm dim}}}        %%several definitions
\def\End{\hbox{{\rm End}}} 
\def\Tr{\hbox{{\rm Tr}}} 
\def\tr{{\hbox{\rm\small{Tr}}}}                %%Trace
\def\lin{{\hbox{Lin}}}
\def\vol{{\hbox{vol}}}  
\def\Hom{{\hbox{Hom}}}
\def\div{{\hbox{div}}} 
\def\rank{{\hbox{rank}}}
\def\Ad{{\hbox{Ad}}}
\def\ad{{\hbox{ad}}}
\def\CoAd{{\hbox{CoAd}}}
\def\coad{{\hbox{coad}}}                           
\def\Rea{\hbox{Re}}                     %  parte real
\def\id{{\hbox{id}}}                    %  la identidad
\def\Id{{\hbox{Id}}}
\def\Int{{\hbox{Int}}}
\def\Ext{{\hbox{Ext}}}
\def\Aut{{\hbox{Aut}}}
\def\Card{{\hbox{Card}}}
\def\SODE{{\small{SODE }}}

%Si no se puede utilizar el fichero mssymb, los fuentes AmS TeX se
%pueden cargar a mano (por ejemplo) con las lineas siguientes
%\newfont{\got}{eufm10 scaled\magstep1}
%\newfont{\field}{msym10 scaled\magstep1}

\newtheorem{theorem}{Theorem}
\newtheorem{corollary}{Corollary}
\newtheorem{proposition}{Proposition}
\newtheorem{definition}{Definition}
\newtheorem{lemma}{Lemma}
\newtheorem{example}{Example}
\newtheorem{remark}{Remark}

\def\Eq#1{{\begin{equation} #1 \end{equation}}}
\def\R{\Bbb R}
\def\C{\Bbb C}
\def\Z{\Bbb Z}
\def\d{\partial}

\def\la#1{\lambda_{#1}}
\def\teet#1#2{\theta [\eta _{#1}] (#2)}
\def\tede#1{\theta [\delta](#1)}
\def\N{{\frak N}}
\def\Wei{\wp}
\def\Hil{{\cal H}}

\font\frak=eufm10 scaled\magstep1

\def\bra#1{\langle#1|}
\def\ket#1{|#1\rangle}
\def\goth #1{\hbox{{\frak #1}}}
\def\<#1>{\langle#1\rangle}
\def\cotg{\mathop{\rm cotg}\nolimits}
\def\cotanh{\mathop{\rm cotanh}\nolimits}
\def\arctanh{\mathop{\rm arctanh}\nolimits}
\def\wt{\widetilde}
\def\const{\hbox{const}}
\def\grad{\mathop{\rm grad}\nolimits}
\def\Div{\mathop{\rm div}\nolimits}
\def\braket#1#2{\langle#1|#2\rangle}
\def\Erf{\mathop{\rm Erf}\nolimits}

\def \Hil {\mathcal{H}}
\def \F {\mathcal{F}}
\def \u{\mathfrak{u}}
\def \S{\mathcal{S}}
\def \O{\mathcal{O}}
\def \D {\mathcal{D}}
\def \Tr {\mathrm{Tr}}
\def \R {\mathbb{R}}
\def \C {\mathbb{C}}
\def \PC {\mathbb{CP}}
\def \PH {\mathcal{PH}}
\def \CP {\mathbb{CP}}

\centerline{\Large \bf Geometrization of Quantum Mechanics}

\vskip 1cm

\begin{center}
Jos\' e F. Cari\~nena$^{\dagger}$, Jes\' us Clemente-Gallardo$^{\star}$
and Giuseppe Marmo$^{\ddagger}$\\
$^{\dagger}$Departamento de  F\'{\i}sica Te\'orica, Universidad de Zaragoza,\\
\smallskip
50009 Zaragoza, Spain.\\
\smallskip

$^{\star}$Instituto de Biocomputaci\'on y F\'{\i}sica
de los Sistemas complejos, \\ Universidad de
Zaragoza, 50009 Zaragoza, Spain.

\smallskip
$^{\ddagger}$Dipartimento di Scienze Fisiche,
Universit\`a Federico II di Napoli\\
 {\small and}\\
INFN, Sezione di Napoli,
Complesso Universitario di Monte Sant'Angelo\\
Via Cintia, 80126 Napoli, Italy\\

\end{center}

\vskip 1cm

\begin{abstract}   
We show that it is possible to represent various descriptions of Quantum
Mechanics in geometrical terms. In particular we start with the space of
observables and use the momentum map associated with the unitary group to
provide an unified geometrical description for the different pictures of
Quantum Mechanics. This construction provides an alternative to the usual GNS
construction for pure states.

\end{abstract}


\section{Introduction}

\qquad The description of  any physical system, either classical or quantum, requires
the identification of a space of states $\S$, a space of observables $\O$, and a
pairing between both spaces with values in the field of real numbers
$\R$.  This pairing may be  interpreted as the measurement process.

Usually in classical mechanics the space of states is given the structure of a
real differentiable manifold, the space of observables is identified with the
space of functions on the manifold and the pairing is the evaluation of a
function at the state.  Alternative approaches start with the associative and
commutative algebra of functions (observables) and identify states as
algebra-homomorphisms from the algebra of observables to that of real numbers.
This approach has been advocated by  Alexander Vinogradov in several papers and an
account can be found in \cite{MMTrieste}.

For quantum systems, following Schr\"odinger and Dirac, the primary object is the
space of states $\S$. It is usually identified with a separable complex Hilbert space
$\Hil$ \cite{Dirac:36}. Observables are a derived object and they are identified with
self-adjoint operators acting on $\Hil$ (domain problems are not important at
this stage). The pairing is provided by the expectation values of observables
at elements of $\Hil$. 

An alternative approach, which is a development of
Heisenberg description of quantum mechanics in terms of `infinite-dimensional
matrices', mainly due to  Segal \cite{Segal:1947},  Haag and Kastler
\cite{HaagKast:1964}, starts 
with a 
$\C^{*}$-algebra and the  observables (real dynamical variables) are given by
its self-adjoint elements. States are then constructed as linear functionals on
the algebra with further requirements to be able to interpret their evaluation
on observables in terms of measurement processes. 

In summary, it appears that while classical mechanics has a highly geometric
content in its formulation, quantum mechanics is to be formulated mainly in
terms of (linear) algebraic structures. While the relevant group of
transformations of classical mechanics is hence a subgroup of the
diffeomorphism group of the manifold of states (in the spirit of Klein's
Erlangen Programme), the relevant group of quantum mechanics is a subgroup of
linear transformations. However, it is commonly accepted that classical
mechanics must be a suitable limit of quantum mechanics (this is usually known
as the `correspondence principle'). Therefore, it is reasonable to expect
that a better understanding of this `limit procedure' may be achieved if we
elaborate a `geometrization' of quantum mechanics in such  a way that we might go
beyond the group of linear transformations. Attempts at  a geometrical
formulation of Quantum Mechanics have been made in the past \cite{cirelli1,
cirelli2, cirelli3, bloch, heslot, rowe, field, brody, strocchi, ashtekar,
manko}, but approaching the problem from a different perspective, namely by 
starting from the Hilbert space of states and trying to identify the necessary
geometrical structures arising from the Hermitian structure.

The aim of this paper is  to outline another way to achieve the geometrization.
We shall follow an approach
close to the $\C^*$--algebraic one  we have mentioned earlier. To avoid technicalities
which may obscure the mathematical geometric structures we are going to
introduce, we shall limit ourselves to finite-level quantum systems. We hope
that in this way the geometrical structures will emerge more neatly.

Our primary object will be the family of Hermitian operators (observables) in $\Hil$,
which is in a one-to-one correspondence with the real Lie algebra $\u(\Hil)$ of the unitary group
$U(\Hil)$. The
real vector space $\u(\Hil)$ carries not only the structure of a Lie algebra, but 
also the structure
of a Jordan algebra \cite{Jordan:1934}, and these two structures are compatible in a sense that
will be clear later. 

A `geometrization' of the Lie algebra structure on the space of observables
is easily achieved by replacing the Lie commutator with a linear Poisson tensor
defined on the dual space ${\goth u}^*(\Hil)$ of the Lie algebra ${\goth
  u}(\Hil)$.  In a similar manner we also
geometrize the Jordan algebra structure defining the corresponding tensor on
the same dual vector space.
 With this replacement we are not
restricted to just linear transformations anymore, but both ${\goth u}(\Hil)$
and ${\goth u}^*(\Hil)$ are considered as $n^2$-dimensional manifolds. 

Starting with the Poisson manifold structure on
this space we may look for its symplectic realizations. In particular, a
symplectic realization by means of a symplectic vector space replaces the more
familiar Gelfand-Nairmark-Segal (GNS) construction of the Hilbert space
associated with the $\C^{*}$-algebra of dynamical variables starting from
 a pure state. 

The  symplectic vector space, equipped with an appropriate complex structure,
will give rise to a Hilbert space $\Hil$. The symplectic action of the 
corresponding unitary group provides us with 
 a momentum map $\mu:\Hil\to \u^*(\Hil)$ 
which defines the sought symplectic realization of the space of
observables we were looking for.   We relate also the space of pure states with
the complex projective space which turns out to be symplectomorphic with minimal
symplectic orbits of the corresponding coadjoint action (of the unitary group)
on the space $\u^*(\Hil)$.

\section{The space of observables}

Following the algebraic approach advocated by Segal \cite{Segal:1947} and
Haag and Kastler \cite{HaagKast:1964}, we consider the space of observables as
the collection of all the self-adjoint elements of a $\C^{*}$-algebra with
identity element. The set of states $\S$ is then the collection of all positive
real valued  linear functionals $\Phi$, normalized by the condition 
\begin{equation}
  \label{eq:ioden}
  \Phi(\mathbb{I})=1.
\end{equation}
The pairing of both sets is provided by evaluation of the state, i.e. taking
the expectation value of the observable 
when the state of the system is $\Phi$.

\subsection{The algebraic structures}
There is also a well-known result  that states that every
$\C^{*}$-algebra with identity can be realized as the set of all bounded
operators acting on some Hilbert space $\Hil$ (see \cite{Emch} for details).
Having decided to deal with
finite-level quantum systems, we shall identify the set of observables with the
space of Hermitian operators (i.e. those operators satisfying $A^{\dagger }=A$,
condition which is now to be written as $A^*=A$ with the $\C^*$--involution).
They define a real vector space isomorphic with the algebra $\u(\Hil)$, which corresponds
to the operators which are anti-Hermitian and which, in the case of finite
dimensional systems, define the Lie algebra $\u(n)$, $n$ being the dimension of
the Hilbert space.

The multiplication by the imaginary unit $i$ establishes a vector space isomorphism
with the real vector space of the Lie algebra of the unitary group $\u(\Hil)$
\begin{equation}
  \label{eq:isomor}
 \alpha: \O\to \u(\Hil)\quad  \alpha(A)= iA.
\end{equation}

 As a
result, as we are considering the case of a finite dimensional Hilbert space it
is immediate to prove that the space $\O$ may be  identified also with the space
$\u^*(\Hil)$.  We summarize this result in the form:

\begin{lemma}
The space of observables $\O$ becomes endowed with a Lie algebra structure,
isomorphic to the natural structure on $\u(\Hil)$  by defining
\begin{equation}
  \label{eq:Liealgebra}
  [A, B]_{-}=\alpha^{-1}(\alpha(A)\alpha(B)-\alpha(B)\alpha(A))=-i[\alpha(A),\alpha(B)],
\end{equation}
along with a real scalar product that is invariant under the adjoint
representation, given by 
\begin{equation}
  \label{eq:scalar}
  \langle A, B\rangle =\frac 12 \Tr (AB)\,.
\end{equation}  
\end{lemma}

Another isomorphism can be defined identifying the vector spaces $\u(\Hil)$
and $\u^*(\Hil)$, according to the pairing given by the Killing-Cartan form
\begin{equation}
  \label{eq:pairing}
  \xi (A)=\frac i2 \mathrm{Tr}(\xi A) \quad \xi \in  \u^*(\Hil), A\in \u(\Hil)
\end{equation}

The space of Hermitian operators carries also another binary product (usually
called Jordan product) defined by 

\begin{equation}
  \label{eq:jordan}
  A\circ B=\frac 12 (AB+BA)=[A, B]_+ .
\end{equation}
 
Let us recall, for completeness, the definition of Jordan algebra:
\begin{definition}
  A non-associative algebra $(\mathcal{A}, \cdot)$ is called a {\bf Jordan algebra} if
  and only if, the operation is commutative and given two arbitrary elements
  $x,y\in \mathcal{A}$, $    (xy)x^2=x(yx^2)$
\end{definition}

With this definition we can conclude
\begin{lemma}
 $(\O, \circ)$ is  a Jordan algebra. 
\end{lemma}
\begin{proof}
The commutativity is trivial to prove. The second condition 
 follows from the associativity of the original product:
$$
[[x,y]_+,x^2]_+= 2(xy+yx)x^2+2x^2(xy+yx)=2(xyx^2+yx^3+x^3y+x^2yx)
$$
$$
[x,[y,x^2]_+]_+=2x(yx^2+x^2y)+(yx^2+x^2y)x=2(xyx^2+x^3y+yx^3+x^2yx)
$$
\end{proof}

\begin{proposition}
The scalar product \eqref{eq:scalar} is also invariant with respect to this new
product, and we have:
\begin{equation}
  \label{eq:compa}
  \langle [A,B]_{-}, C\rangle =\langle A, [B,C]_{-}\rangle \,, \qquad  \langle [A,B]_{+}, C\rangle =\langle A, [B,C]_{+}\rangle\,.
\end{equation}

Moreover we also have the compatibility relation
\begin{equation}
  \label{eq:compatb}
  [A, B\circ C]_{-}=[A,B]_{-}\circ C+B\circ [A, C]_{-}\,
\end{equation}  
i.e. ad$_-$ is a derivation of the Jordan algebra.
\end{proposition}
\begin{proof}
These properties follow directly from the definitions.   
\end{proof}

From \cite{landsman,Emch}, we define further
\begin{definition}
  A Lie algebra $(\mathcal{A},[\cdot, \cdot ]_-)$ is called a {\bf Jordan-Lie
    algebra} if it admits a Jordan algebra structure $\circ$ 
  satisfying (\ref{eq:compatb}) and the associator identity, for some $\hbar \in \R$ 
  \begin{equation}
    \label{eq:associator}
    (A\circ B)\circ C-A\circ (B\circ C)=\frac 14 \hbar [ [ A, C]_- , B]_- \quad \forall A,B,C\in \mathcal{A}
  \end{equation}
\end{definition}

\begin{lemma}
  The space of observables, with the operations defined above, becomes also a
  Jordan-Lie algebra.
\end{lemma}
\begin{proof}
  See \cite{Emch} for details.
\end{proof}

\subsection{The geometrical structures}
A specific way to geometrize the Lie algebra structure of $\u(\Hil)$ is to
associate with it a linear of Poisson tensor on the dual vector space
$\u^*(\Hil)$ as follows. As
$\Hil$ is assumed to be finite-dimensional, we can
identify $\u(\Hil)$ with the space of real valued linear functions on its dual
space, i.e. $\u(\Hil)\sim \mathrm{Lin}(\u^*(\Hil), \R)$, and we set, for any pair of
linear functions on $\u^*(\Hil)$ defined by the  two elements $u,v\in \u(\Hil)$:
\begin{equation}
  \label{eq:Lie}
  \{ \hat u,\hat v\}=\widehat{[u,v]}\,,
\end{equation}
where the commutator on the right hand side is computed by thinking of $u,v$ as
elements of the Lie algebra  $\u(\Hil)$, and the left hand side is to be read
as a linear function on $\u^*(\Hil)$. We will use the `hat' to denote the
elements of $\u(\Hil)$ seen as linear functions on the dual $\u^*(\Hil)$.
 We are implicitly using here the property that
the vector space $\u(\Hil)$ is isomorphic to its bidual, which holds for vector
spaces which are reflexive, in particular finite dimensional ones. Then, we have: 
\begin{proposition}
  Let $\O$ be the space of observables of a finite level quantum system. Then,
  $\O^*$ (and $\O$ itself) can be endowed with a Poisson structure.
\end{proposition}

Having replaced the Lie algebra structure with the Poisson tensor associated
with the Poisson bracket on $\u^*(\Hil)$, we are now able to perform also
nonlinear transformations on the Poisson manifold. In
this sense we speak of the `geometrization' of the algebra structure of the
vector space $\u(\Hil)$. 

If we denote by $\hat A$ and $\hat B$ the linear functions on $\u^*(\Hil)$
corresponding to elements $A,B\in \u(\Hil)$, we can define the Poisson bivector $\Lambda$
as:
\begin{equation}
  \label{eq:poissontensor}
  \Lambda(d\hat A, d\hat B)(\xi)=\{ \hat A, \hat B\} (\xi)=\xi([A,B])=\frac i2 \Tr\xi
 (AB-BA)\,, \quad \xi \in \u^*(\Hil)\, ,
\end{equation}
where we used the relation (\ref{eq:pairing}).

Hence we recover the well-known Kirillov-Konstant-Souriau Poisson
tensor on the dual of any Lie algebra.

By using a similar procedure we may also `geometrize' the Jordan algebra
structure on the space of observables. Again we set:
\begin{equation}
  \label{eq:jordantensor}
  \mathcal{R}(\xi)(d\hat A, d\hat B)=\xi([A,B]_+)=
\frac 12 \Tr \xi (AB+BA)\,, \quad \xi \in \u^*(\Hil)\, ,
\end{equation}
where use is made of (\ref{eq:pairing}).

These two tensor fields can be put together to form a complex vector field:
\begin{equation}
  \label{eq:assocuativetensor}
  (\mathcal{R}+i\Lambda)(\xi)(d\hat A, d\hat B)=2(\widehat{AB})(\xi )=\xi (AB) =
\Tr(\xi AB)\,, \quad \xi \in \u^*(\Hil)\,.
\end{equation}
By using this tensor field we can define a $*$--product in the form
$$
(\hat A* \hat B)(\xi)=\xi (AB)=(\mathcal{R}+i\Lambda)(\xi)(d\hat A, d\hat B).
$$

In this context, the compatibility condition of the Lie and the Jordan
structures can be simply stated by saying

\begin{proposition}
The Hamiltonian vector fields
associated with observables (i.e. linear functions on $\u^*(\Hil)$) are
infinitesimal symmetries for the tensor field $\mathcal{R}$ associated with the
Jordan structure, and therefore derivations for the $*$--product.  
\end{proposition}
\begin{proof}
  It is a direct consequence of the compatibility between both brackets,
  summarized in (\ref{eq:compatb}).
\end{proof}

\subsection{Distributions}
In this section we will summarize very quickly the results presented in
\cite{Grabowski:2005,Grabowski:2006}, regarding the definition of suitable
distributions 
defined on $\u(\Hil)$ which correspond to the geometric objects defined
above.  By using the identification of the real vector space $\u(\Hil)$ with its tangent
space at each point, we may define a $(1,1)$-tensor field $\hat J$ 
in $\u(\Hil)$ by setting
\begin{equation}
\hat J_\xi (A)=[A,\xi]_-=\Lambda_\xi (d\hat A)\,,\quad A\in \u(\Hil).
\end{equation}
Similarly we can write:
\begin{equation}
\hat {\mathcal{R}}_\xi (A)=[A,\xi]_+=\mathcal{R}_\xi (d\hat A)\,,\quad A\in \u(\Hil).
\end{equation}

These two tensors allow us to define two important distributions:
\begin{equation}
  \label{eq:distribu}
  D_\Lambda =\mathrm{Image}(\hat J) \,,\quad  D_{\mathcal{R}} =\mathrm{Image}(\hat {\mathcal{R}}) 
\end{equation}

Considering these two objects we find that both tensors commute among
themselves:
\begin{equation*}
  \hat J\circ {\hat {\mathcal{R}}}(A)=\hat {\mathcal{R}}\circ \hat J(A)=\frac 12[A, \xi^2]_-.
\end{equation*}

Both distributions can be combined into two new ones: $D_0=D_{\mathcal{R}}\cap D_\Lambda
$ and $D_1=D_{\mathcal{R}}+D_\Lambda$ .

\begin{proposition}
The distributions $D_{0}, D_{1}, D_{\mathcal{R}}$ and $ D_{\Lambda }$ satisfy (\cite{Grabowski:2005}):
\begin{itemize}
\item $D_\Lambda$ is an involutive distribution providing a generalized foliation
  whose leaves are the symplectic manifolds associated with the Poisson tensor
  $\Lambda$. 
\item The distribution $D_{\mathcal{R}}$ is not involutive.
\item The distribution $D_0$ is involutive. The corresponding foliation is
  invariant under the  adjoint representation and each leaf is a K\" ahler
  manifold.
\item The distribution $D_1$ is involutive and the corresponding leaves of its
  associated foliation are orbits of the $\mathrm{GL}(\Hil)$ action defined as
  \begin{equation}
    \label{eq:GLaction}
    \mathrm{GL}(\Hil)\times \u(\Hil)\to \u(\Hil)\,, \quad (T, A)\mapsto TAT^\dagger .
  \end{equation}
\end{itemize}  
\end{proposition}

\begin{remark}
Our `geometrization' carries along the possibility of performing nonlinear
transformations because we have replaced the algebraic structures on the linear
space $\u(\Hil)$ with tensorial objects on the manifold $u^ *(\Hil)$. It should
be remarked, however, that now on $\u^*(\Hil)$ 
we have the possibility of two different products on linear functions:
\begin{itemize}
\item the pointwise product $(\hat A\cdot \hat B)(\xi)=\hat A(\xi)\hat B(\xi)$, which
  gives a quadratic function out of two linear ones and
\item a non-local product $(\hat A\star \hat B)(\xi)=\widehat{AB}(\xi)$. In this case we
  obtain a linear function as the product of other two linear ones, but in general it
  will be a complex valued function even if the factors were real ones. This
  result has to do with the fact that the product of two Hermitian operators is
  not Hermitian and therefore it gives rise to real and imaginary parts.
\end{itemize}
\end{remark}

\begin{example}
  At this point it may be adequate to give a simple example of the objects
  introduced so far. Let us consider the Lie algebra ${\goth{su}}(2)$ of $2\times 2$ Hermitian
  matrices corresponding to a spin $1/2$ physical system. We introduce an
  orthonormal basis with respect to the scalar product \eqref{eq:scalar}. We
  set thus:
  \begin{equation*}
    U=\begin{pmatrix}
1& 0\\
0&1
\end{pmatrix},
\quad  
    X=\begin{pmatrix}
0& 1\\
1&0
\end{pmatrix},
\quad 
    Y=\begin{pmatrix}
0&- i\\
i&0
\end{pmatrix},
\quad
    Z=\begin{pmatrix}
1& 0\\
0&-1
\end{pmatrix},
\quad \,,
  \end{equation*}
and also the associated linear functions
$$
\hat X=x,\quad \hat Y=y, \quad \hat Z=z, \quad \hat U=u 
$$
where the functions are to be understood as $z(A)=\frac i2 \Tr(ZA)$ ,and so on,
for any $A\in u(2)$. In these coordinates, the Poisson tensor field is given by
\begin{equation*}
  \Lambda=2\left (x\frac\partial{\partial y}\land  \frac \partial{\partial z}+y\frac\partial{\partial z}\land \frac \partial{\partial
      x}+z\frac\partial{\partial x}\land \frac \partial{\partial y}\right ) ,
\end{equation*}
while the tensor associated to the Jordan structure becomes:
\begin{eqnarray}
  \mathcal{R}&=&2\frac \partial {\partial u}\otimes_s \left ( x\frac \partial{\partial x}+y\frac \partial{\partial y}+z\frac \partial{\partial z}\right
  )\cr&+&2u \left ( \frac\partial{\partial u}\otimes_s\frac\partial{\partial u}+ \frac\partial{\partial
      x}\otimes_s\frac\partial{\partial x}+  \frac\partial{\partial y}\otimes_s\frac\partial{\partial
      y}+  \frac\partial{\partial z}\otimes_s\frac\partial{\partial z} \right ),\nonumber
\end{eqnarray}
where $\otimes_s$ stands for the symmetrized tensor product.

It is immediately  seen that $\mathcal{R}$ is invariant under the rotation vector fields
provided by the linear Hamiltonian functions with respect to the Poisson
tensor $\Lambda$. We can even consider the non-local product, for instance we get
$$
 \hat Z\star \hat Y=-i\hat X ,\quad \hat X\star \hat Y=i\hat Z ,\quad \hat Z \star\hat X=i\hat Y.
$$

It is also easy to see that the Hamiltonian vector fields associated with
linear functions provide derivations both for the pointwise product and for the
non-local product.  Thus, the associated equations of motion  do not carry a
quantum or a classical label, it is the product what distinguishes the
commutative or the non-commutative nature of the space along with the locality
or non-locality of the operation. And therefore distinguishes Classical from
Quantum Mechanics.
\end{example}

\subsection{Dynamics}

It is now possible to write equations of motion on the phase space of
observables. In the pure Heisenberg picture it is defined as
$$
\frac d{dt}A=\frac 1{\hbar} [H,A]_-\,.
$$

By using the `geometrization', i.e. by thinking in terms of  the dual space
$\u^{*}(\Hil)$, we find  
$$
\frac d{dt}\widehat A=\frac 1{\hbar} \, \{\widehat  H, \widehat A\}\, .
$$

As there are different algebra structures on $\u^{*}(\Hil)$ it is important to
study the compatibility of differential equations with such algebra structures.
\begin{lemma}
The linear differential equations which preserve both products correspond to
the infinitesimal generators of unitary transformations.
\end{lemma}

\subsection{States}

States are identified as the elements of the convex body 
$
\mathcal{S}=\{\phi\in\mathcal{A}^*\mid \phi(A^*A)\geq 0\,, \forall A\in
 \mathcal{A}; \phi(\uno)=1\}\,.
$

It is not difficult to show in the finite-dimensional case, for the
infinite-dimensional case it is a theorem by Gleason, that any state can be
written in the form  
$
\phi(A)={\rm Tr\,}\rho_\phi A\,.
$

Moreover, $\rho_\phi$ is a nonegatively defined operator of ${\goth{gl}}(\Hil)$,
i.e. those  $\rho_\phi\in {\goth{gl}}(\Hil)$ which can be written in the form 
$\rho_\phi=T^\dag T$ for some $T\in {\goth{gl}}(\Hil)$ and, in addition, satisfy 
${\rm Tr\,}\rho_\phi=1$. Thus, $\mathcal{S}$ is a convex body in the
 affine hyperplane in ${\goth{u}}^*(\Hil)$, determined by the equation
 ${\rm Tr\,}\rho_\phi=1$.
The tangent space to this 
affine hyperplane at a point is therefore identified with
the space of traceless Hermitian operators, and it is in a one-to-one correspondence with the
 Lie algebra of the group $SU(\Hil)$. 
\section{Symplectic realizations of the Poisson manifold ${\biggoth{u}}^*$} 
\subsection{The representation in the finite dimensional case}
In the general case, to reconstruct a Hilbert space representation from the
`algebraic description'  of quantum systems, along with its `geometrization',
we can use the Gelfand-Naimark-Segal (GNS) construction.  This requires the
introduction of a state (a functional on the algebra of observables which
satisfies certain conditions) and the definition of an adapted Hilbert space
where the algebra of observables can be represented.

In this finite-dimensional  situation there is a simpler alternative. 
We may use techniques from differential geometry of Hamiltonian systems to
 construct a representation on a Hilbert space. Specifically, 
we search for a symplectic realization of the Poisson manifold by means of a symplectic 
vector space (a classical Jordan-Schwinger map)
\begin{definition}
  A {\bf symplectic realization} of  a Poisson manifold $(N,\{\cdot, \cdot\} )$ 
is a Poisson map $\Phi:M\to  N$, where  $(M,\omega)$ is a
symplectic manifold. When $M$ is a symplectic vector space we have a special
situation and $\Phi$ is called a {\bf classical Jordan-Schwinger map} \cite{liejordan}.
\end{definition}

\subsubsection{The K\"ahler space}
The first step is to consider the dimension of the algebra of observables.
From the dimension of ${\goth{u}}(\Hil)$, say $n^2$-dimensional, we may 
consider the action of the unitary group $U(n)$ on the vector space
 $\mathbb{R}^{2n}$, endowed with a K\"ahler structure, a triple
 $(\omega,g,J)$. Here $\omega$ is an exact symplectic structure, $g$ is an
 Euclidean structure and $J$ the associated complex structure. 
We consider what is known as the ``defining representation''
of the group $U(n)$. 

The symplectic structure admits a potential 1-form which is invariant under the
action and then  the associated momentum map provides us with the symplectic
realization we are  searching for (see \cite{BCG:1991}). The momentum map,  $\mu:
\mathbb{R}^{2n}\to {\goth{u}}^*(n)$, is equivariant with respect to the fundamental action of
$U(n)$ on $\mathbb{C}^n\approx \mathbb{R}^{2n}$ and the coadjoint action on  
${\goth{u}}^*(n)$.

If we select an orthonormal basis $\{e_1,\ldots,e_n\}$ for $\mathbb{C}^n$, 
we may define coordinates by setting 
$\braket{e_k}{\psi}=z_k(\psi)=(q_k+i\, p_k)(\psi)$, and we have used Dirac's
 notation for bras and kets.

\subsubsection{The geometric structures}
One can see (\cite{Grabowski:2005}) that in these
coordinates we have a  contravariant version of the Euclidean structure given by 
$
G=\sum_{k=1}^n\left(\pd{}{q_k}\otimes \pd{}{q_k}+\pd{}{p_k}\otimes \pd{}{p_k}\right)
$
and the Poisson tensor
$
\Omega=\sum_{k=1}^n\left(\pd{}{q_k}\land \pd{}{p_k}\right)
$
while the complex structure has the form 
$J=\sum_{k=1}^n\left(\pd{}{p_k}\otimes d{q_k}+\pd{}{q_k}\otimes d{p_k}\right)
$.

In terms of complex coordinates the Hermitian structure has the form 
$h=\sum_{k=1}^n d\bar z_k\otimes  dz_k$. The corresponding contravariant form is given by 
$$G+i\,\Omega=\sum_{k=1}^n\left(\pd{}{q_k}-i\pd{}{q_k}\right)\otimes
\left(\pd{}{q_k}+i\pd{}{q_k}\right)=4\,  \sum_{k=1}^n \pd{}{z_k}\otimes
\pd{}{\bar z_k}
$$

We may define binary products on functions by setting 
$$
\begin{array}{rcl} \{f_1,f_2\}&=&{\displaystyle\sum_{k=1}^n\left(\pd
  {f_1}{q_k}\,\pd{f_2}{p_k}-
\pd {f_1}{p_k}\,\pd{f_2}{q_k}\right)}\cr
\{f_1,f_2\}_+&=&{\displaystyle\sum_{k=1}^n\left(\pd
  {f_1}{q_k}\,\pd{f_2}{p_k}+
\pd {f_1}{p_k}\,\pd{f_2}{q_k}\right)}\cr
\braket{f_1}{f_2}&=&4\ {\displaystyle\sum_{k=1}^n\pd{f_1}{z_k}\,\pd{f_2}{\bar z_k}}
\end{array}
$$ 

\subsection{The representation of $\O$}
Now we consider the corresponding representation of the algebra of
observables. First we can consider the following trivial lemma: 
\begin{lemma}
Every complex linear operator $A\in {\goth{gl}}(n,\mathbb{C})$ 
defines a quadratic function
$f_A(\psi)=\frac 12 \braket{\psi}{A\psi}$. The function is real if and only if $A$ is
Hermitian, $A=A^\dag$.   
\end{lemma}

Having introduced the relevant geometrical structures associated with
 the K\" ahler structure, we may now proceed to establish the connection
 between the description on the space of observables, ${\goth{u}}(\Hil)$,
 with the description on the Hilbert space $\Hil\equiv \mathbb{C}^n$.

 \begin{proposition}
   
 The main properties of the momentum map, 
$$
\mu: \Hil\to {\goth{u}}^*(\Hil)\,,\qquad \mu:\ket\psi\mapsto \rho_\psi=\ket\psi\bra\psi\,,
$$ 
are the following:

\begin{itemize}
\item $\mu^*(\widehat A)=f_A$
\item  $\mu^*(\{\widehat A,\widehat B\})=\{f_A,f_B\}$
\item $\mu^*(\mathcal{R}(d\widehat A, d\widehat B))=G(\mu^*( d\widehat A), \mu^*(
  d\widehat B))$ 
\end{itemize}
\end{proposition}
\begin{proof}
  Direct verification.
\end{proof}

Thus we conclude that the  geometrization of the observables on the Hilbert
space corresponds to the study of the associated quadratic functions.

\subsection{Dynamics}
It is now possible to write the dynamics on the Hilbert space, Schr\"odinger equation,
by writing 
$
i\, \hbar \frac d{dt}\ket\psi=H\ket \psi
$
or equivalently,
$
 \hbar\,\frac d{dt}\,f_A=\{f_H,f_A\}\,.
$

This last equation is $\mu$-related with the Heisenberg equation on the space of observables.

\subsection{Eigenvalues and eigenstates}
In the `geometrized' Hilbert space description we have to recover now the
description of the `eigenvectors' and `eigenvalues'. 

To this aim is appropriate to introduce expectation values associated with
Hermitian operators, $A=A^\dagger$: 
$$
e_A(\psi)=\frac{\braket{\psi}{A\psi}}{\braket{\psi}{\psi}}\,.
$$

We find that:

\begin{enumerate}
\item Critical points of $de_A$  correspond to the eigenvectors of $A$.
\item Values of $e_A$ at critical points are the corresponding eigenvalues of $A$.
\end{enumerate}

\begin{remark}
   Critical points of $de_A$ coincide with the critical
 points of the corresponding Hamiltonian vector field $\Omega(de_A)$ or 
of the corresponding
gradient vector field $G(de_A)$.
\end{remark}

\begin{remark}
$$G(de_A,de_A)= \frac{\braket{\psi}{A^2\psi}}{\braket{\psi}{\psi}}-
\frac{\braket{\psi}{A\psi}}{\braket{\psi}{\psi}}
\frac{\braket{\psi}{A\psi}}{\braket{\psi}{\psi}}
\,.
$$
i.e. it represents the dispersion of the mean value of the observable
corresponding to the operator  $A$
 in the state $\ket\psi$, therefore the square of the Hamiltonian vector
 field associated with $A$ is strictly related to the `uncertainty' in the
 measurement of $A$ in the state $\ket\psi$.  
\end{remark}

\begin{remark}
The GNS construction associated with a pure state would provide us with
 a complex Hilbert space of dimension $n$ if we start with
 $n\times n $-Hermitian matrices.  Therefore, it may be considered as a way to
 construct a ``symplectic realization'' of the algebraic structures on
 $\u^*(\Hil)$ (which also realizes the Jordan algebra). 
\end{remark}

\section{Conclusions}

In this paper we have presented a way of providing a geometrization of
Quantum Mechanics starting from the set of observables of our quantum
system. To close the circle, we may try to connect our results with the
usual geometrical constructions which take the space of states as starting
point. 

The probabilistic interpretation of quantum mechanics requires the states to
 be normalized and, moreover, because normalized vectors are
 identified with probability amplitudes, the physically relevant
 probability densities are invariant under multiplication of vectors by a
 phase. In conclusion, physical states should be identified not with 
vectors in the Hilbert space but rather with  points in the associated complex
 projective space. We thus define the principal fibration 
$\mathbb{C}_0\to{\Hil}\,\--\,\{0\}\to \mathbb{P}\Hil\,,$
where $\mathbb{C}_0=\mathbb{C}-\{0\}$ is the set of nonzero complex
 numbers and ${\Hil}-\{0\}$ that of nonzero vectors. $\mathbb{P}\Hil\,,$ denotes
 the complex projective space associated with $\Hil$.

From the manifold point of view the quotient $\mathbb{P}\Hil$  is completely
described by the  involutive distribution generated by $\Delta$, the infinitesimal generator 
of dilations, and by $J(\Delta)$, the infinitesimal generator of
 multiplication by a phase. They provide enough information to describe the
 points of the projective space. 
The relevant tensors $G$ and $\Omega$ introduced in the previous section, however,
are not projectable with respect to the fibration above.
 To turn them into projectable tensors we can multiply them
by a conformal  factor $\braket{\psi}{\psi}$ and in this way we get projectable tensor
 fields. However,  $\braket{\psi}{\psi}\, \Omega$ does not define a bracket
 satisfying the Jacobi identity. As a matter of fact, it gives rise to a Jacobi
 bracket. Therefore, the Poisson bracket we obtain on the complex projective
 space has to be considered as a reduction of a Jacobi bracket on $\Hil$
 instead of a Poisson bracket on $\Hil$. Further details can be found in
 \cite{manko}.

To close the circle, we notice that the complex projective space may be
identified with minimal symplectic orbits on $\u^*(\Hil)$ (see
\cite{Grabowski:2005}).  The Schr\"odinger
equations of motion on the Hilbert space project onto  von Neumann equations of
motion on the space of pure states (rank one projectors). See \cite{manko2}.

With this observation we conclude by saying that (all) various formalisms to
describe quantum systems may be `geometrized' and these `geometrizations' rely
on the consideration of the momentum map associated with the strongly
Hamiltonian action of the unitary group on the Hilbert space of states.  The
treatment has been  carried out
for finite level quantum systems. However the extension to infinite-dimensional
Hilbert spaces encounters only problems we can call technical, the conceptual
geometrical framework is fully captured by the finite dimensional situation.


\begin{thebibliography}{99}

\bibitem{MMTrieste} 
G. Marmo and G. Morandi, {\it Some Geometry and Topology} in {\bf Low
  dimensional Quantum Field Theory for condensed matter physicists}, Lecture
Notes of ICTP,1--108, Trieste (1992)

\bibitem{Dirac:36} 
P.A.M. Dirac,  {\it The Principles of Quantum Mechanics}, Clarendon Press, Oxford, 2nd
edition  (1936).

\bibitem{Segal:1947}
E.~Segal,
{\it ``Postulates for general quantum mechanics''}, 
Annals of Math. {\bf 48}, 930--948 (1947).

\bibitem{HaagKast:1964}
R.~Haag and D.~Kastler,
{\it ``An algebraic approach to quantum field theory''},
J. Math. Phys. {\bf 5}, 848--861 (1964).

\bibitem{cirelli1}
R. Cirelli, P. Lanzavecchia and A. Mani\' a, {\it  Normal pure states of the von
  Neumann algebra of bounded operator as K\" ahler manifold}, J. Phys. A:
Math. Gen. {\bf 15}, 3829-3835  (1983)

\bibitem{cirelli2}
R. Cirelli and P. Lanzavecchia, {\it  Hamiltonian vector fields in Quantum
  Mechanics},  Nuovo Cimento B, {\bf 79}, 271--283 (1984)

\bibitem{cirelli3}
M.C. Abbati, R. Cirelli,  P. Lanzavecchia and A. Mani\' a, {\it Pure states of
  general quantum mechanical systems as K\" ahler bundle},
Nuovo Cimento B, {\bf 83}, 43--60 (1984)

\bibitem{bloch}
A. Bloch, {\it An infinite-dimensional Hamiltonian system  on a projective
  Hilbert space}, Trans. AMS {\bf 302}, 787--796 (1987)

\bibitem{heslot}
A. Heslot, {\it Quantum mechanics as a classical theory}, Phys Rev D {\bf 31},
1341--1348 (1985)

\bibitem{rowe}
D.J. Rowe, A. Ryman and G. Rosensteel, {\it Many body quantum mechanics as a
  symplectic dynamical system}, Phys Rev A {\bf 22}, 2362-2372  (1980)

\bibitem{field}
T.R. Field and J.S. Anandan, {\it Geometric phases and coherent states},
J. Geom. Phys. {\bf 50}, 56--78 (2004)

\bibitem{brody}
D. Brody and L.P. Hughston, {\it Geometric quantum mechanics},
J. Geom. Phys. {\bf 38}, 19--53  (2001)

\bibitem{strocchi}
F. Strocchi, {\it Complex coordinates and quantum mechanics},
Rev. Mod. Phys. {\bf 38}, 36--40  (1956)

\bibitem{ashtekar}
A. Ashtekar and T.A. Shilling, {\it Geometrical formulation of Quantum
  Mechanics}, in {\bf On Einstein's path}, Ed. A. Harvey, Springer  (1998)

\bibitem{manko}
V. I. Manko, G. Marmo, E.C.G. Sudarshan and F. Zaccaria, {\it The geometry of
  density states}, Rep Math Phys {\bf 55}, 405-422  (2005)


\bibitem{Jordan:1934}
P. Jordan, J. von Neumann and E.P. Wigner,
{\it ``On an algebraic generalization of the quantum mechanical formalism''}, 
Ann. Math. {\bf 35}, 29--64 (1934)

\bibitem{Emch}
G.G. Emch, {\it Foundations of 20th century
  Physics}, North Holland, Amsterdam  (1984)

\bibitem{landsman}
N. P. Landsman, {\it Mathematical Topics Between Classical and Quantum
  Mechanics},  Springer
Mathematical Monographs, Springer, New York, 1998



\bibitem{liejordan}
V.I. Manko, G. Marmo, P. Vitale and F. Zaccaria,
{\it A generalization of the Jordan-Schwinger map: classical version and its
  $q$--deformation}, Int. J. Mod. Phys. A, {\bf 9}, 5541--5561  (1994)


\bibitem{Grabowski:2005}
J. Grabowski, M. Ku\'s and  G. Marmo, 
{\it ``Geometry of quantum systems: density states and entanglement''},
 J. Phys. A:Math. Gen {\bf 38}, 10217-10244  (2005).  %math-ph/0507045

\bibitem{Grabowski:2006}
J. Grabowski, M. Ku\'s  and  G. Marmo, 
{\it ``Symmetry, group actions and entanglement''},
 Open sys. \& Information dyn.  (2006, to appear).  %math-ph/0507045

\bibitem{BCG:1991}
L.J. Boya, J.F. Cari\~nena and  J.M. Gracia-Bond\'{\i}a,
{\it Symplectic structure of the Aharonov-Anandan geometric phase},
{Phys. Lett.\/} A {\bf 161 }, 30--34 (1991).

\bibitem{manko2}
V.I. Man'ko, G. Marmo, E.C.G. Sudarshan and  F. Zaccaria,
{\it ``On the relation between Schr\"odinger and von Neumann equations''}, 
J. Russ. Laser Research {\bf 20}, 421--437  (1999) %quant-ph/0503041

\end{thebibliography}
\end{document}